\documentclass[a4paper,12pt]{article}
\usepackage{jheppub,booktabs}
\usepackage{amsmath,amssymb,bm}
\usepackage{xcolor,colortbl}
\usepackage{cancel}
\usepackage{subfigure}
\usepackage{mathrsfs}
\usepackage{graphicx}
\usepackage{hepunits}
\usepackage[utf8x]{inputenc}

\def\OXY#1#2#3{{\cal O}_{\hspace{-0.1mm} b\hspace{-0.12mm}s_{\hspace{-0.29mm}#1} \hspace{-0.095mm} #3_{\hspace{-0.1mm}#2}}}
\def\CXY#1#2#3{C_{\hspace{-0.1mm} b\hspace{-0.12mm}s_{\hspace{-0.29mm}#1} \hspace{-0.095mm} #3_{\hspace{-0.1mm}#2}}}
\def\CXYSM#1#2{C_{\hspace{-0.1mm} b\hspace{-0.12mm}s_{\hspace{-0.29mm}#1} \hspace{-0.095mm} \ell_{\hspace{-0.1mm}#2}}^{\rm SM}}
\def\CXYNP#1#2#3{C_{\hspace{-0.1mm} b\hspace{-0.12mm}s_{\hspace{-0.29mm}#1} \hspace{-0.095mm} #3_{\hspace{-0.1mm}#2}}^{\rm NP}}

\preprint{CERN-TH-2017-257, MITP/17-095}

\author[a, c]{Adrián Carmona}
\author[b]{and Florian Goertz}
\affiliation[a]{PRISMA Cluster of Excellence \& Mainz Institute for Theoretical Physics, Johannes Gutenberg University, 55099 Mainz, Germany}
\affiliation[b]{Max-Planck-Institut für Kernphysik, Saupfercheckweg 1, 69117 Heidelberg, Germany}
\affiliation[c]{Theoretical Physics Department, CERN, Geneva, Switzerland}
\emailAdd{adrian.carmona@uni-mainz.de}

\emailAdd{fgoertz@mpi-hd.mpg.de}

\title{Recent $\boldsymbol{B}$ Physics Anomalies - a First Hint for Compositeness?}

\abstract{We scrutinize the recently further strengthened hints for new physics in semileptonic
$B$-meson decays, focusing on the 'clean' ratios of branching fractions $R_K$ and $R_{K^\ast}$
and examining to which pattern of new effects they point to. We explore in particular the hardly considered,
yet fully viable, option of new physics in the right-handed electron sector and demonstrate how a recently
proposed framework of leptons in composite Higgs setups naturally solves both the $R_K$ and $R_{K^\ast}$
anomalies via a peculiar structure of new physics effects, predicted by minimality of the model and the scale
of neutrino masses. Finally, we also take into account further observables, such as 
${\cal B}(B_s \to \mu^+\mu^-)$, $\Delta M_{B_s}$,
and angular observables in $B \to K^{\ast} \mu^+ \mu^-$ decays, to arrive at a comprehensive 
picture of the model concerning (semileptonic) $B$ decays. We conclude that 
-- since it is in good agreement with the experimental situation in flavor physics
 and also allows to avoid ultra-light top partners -- the model furnishes a very
promising scenarios of Higgs compositeness in the light of LHC data.}
\date{\today}

\begin{document}
\maketitle

\section{Introduction}

Decays of $B$ mesons offer a promising place to
search for new physics (NP), since in the Standard Model (SM) of Particle
Physics flavor changing neutral processes are strongly suppressed and thus
effects of NP might be sizable in (flavor-changing) $B$ decays. The case is strengthened by 
the fact that the bottom quark is the heaviest down-type quark and resides in the 
same weak doublet as the $t_L$, which due to its large mass is thought to furnish a 
major link to the completion of the SM at smallest distances.

In fact, several anomalies have been found in
$b \to s \ell^+ \ell^-$ transitions, such as the long-standing anomaly
in the angular analysis of the $B \to K^* \mu^+ \mu^-$ decay \cite{Aaij:2013qta,Aaij:2015oid,Wehle:2016yoi,CMS:2017ivg,ATLAS:2017dlm}, 
as well as deficits in the branching fractions $B \to K \mu^+ \mu^-$ \cite{Aaij:2014pli}
and $B_s \to \phi \mu^+ \mu^-$ \cite{Aaij:2015esa}.
While these anomalies might be interpreted as a sign of new physics\footnote{See 
\cite{Descotes-Genon:2013wba,Altmannshofer:2013foa,Beaujean:2013soa,Hurth:2013ssa,Gauld:2013qba,Altmannshofer:2014rta,Descotes-Genon:2015uva,Hurth:2016fbr,Altmannshofer:2017fio} 
for theoretical interpretations in the framework of effective field theory.},
some caution is in order because potentially sizable hadronic uncertainties
are challenging to control.

A much cleaner probe of NP is given by ratios of branching fractions, like
\begin{equation}
\label{eq:RK}
	R_K \equiv \frac{{\cal B}(B^+ \to K^+ \mu^+\mu^-)}{{\cal B}(B^+ \to K^+ e^+e^-)},
\end{equation}
which tests lepton flavor universality (LFU), and where large hadronic uncertainties basically drop
out \cite{Hiller:2003js} (see below). Interestingly, this theoretically very clean observable has also been
measured at LHCb \cite{Aaij:2014ora} and exhibits a sizable ($25 \%$) depletion with respect to the SM 
prediction \cite{Hiller:2003js,Bordone:2016gaq} $\left| R_K^{\rm SM}-1 \right| < 1\%$\,\footnote{This is valid 
above the muon threshold, $q^2 \sim 4 m_\mu^2$.}
in the $q^2 \equiv (p_{\ell^-}+p_{\ell^+})^2 \in [1,6]\,{\rm GeV}^2$ bin, {\it i.e.},
\begin{equation}
	R_K^{[1,6]\, {\rm exp}} \equiv \left. R_K \right|_{q^2 \in [1,6]\,{\rm GeV}^2}^{\rm exp} = 0.745^{+0.090}_{-0.074}\pm0.036\,.
\end{equation}
This corresponds to a deviation of almost $3\,\sigma$.
This intriguing hint for NP was recently strengthened with the measurement
of the ratio
\begin{equation}
	R_{K^\ast} \equiv \frac{{\cal B}(B^0 \to K^\ast \mu^+\mu^-)}{{\cal B}(B^0 \to K^\ast e^+e^-)},
\end{equation}
which was found to be low by a similar amount \cite{Aaij:2017vbb}
\begin{equation}
\begin{split}
	R_{K^\ast}^{[1.1,6]{\rm exp}}= 0.69^{+0.11}_{-0.07}\pm0.05\,,\\
	R_{K^\ast}^{[0.045,1.1]{\rm exp}}= 0.66^{+0.11}_{-0.07}\pm0.03\,,
\end{split}
\end{equation}
where again $| R_{K^\ast}^{[1.1,6]\,\rm{SM}}-1|< 1\%$ and 
$| R_{K^\ast}^{[0.045,1.1]\,\rm{SM}}-1| \lesssim 5\%$, resulting in tensions of $2.5\,\sigma$ and $2.2\,\sigma,$ respectively. 
A naive combination of both results leads to a discrepancy with the SM of about $4\,\sigma$,
employing only observables that are theoretically well under control.\footnote{Note that in the case of $R_K$ this holds also in the presence of NP,
while for $R_{K^\ast}$ it holds within the SM, still allowing for a clean test of the latter.}

In this article, we will focus on these clean observables, and study first the structure 
of NP required to explain the found deviations in an effective field theory (EFT) approach.
We will in particular stress a solution with sizable effects in the right-handed electron sector,
complementary to the common solution which is linked to the left-handed muon sector.
We will then show how a recently proposed composite Higgs model, incorporating a non-trivial, yet minimal,
implementation of the lepton sector \cite{Carmona:2015ena,Carmona:2014iwa} 
can explain the $R_K$ and $R_{K^\ast}$ anomalies {\it simultaneously},
due to the peculiar chirality structure of the involved currents. The setup contains less degrees of freedom than standard
realizations and is very predictive, leading in general to a non-negligible violation of LFU,
while allowing at the same time for a strong suppression of flavor changing neutral currents (FCNCs).\footnote{
Beyond that, in these models the leptonic contribution to the Higgs mass is parametrically
enhanced relative to the quark contribution by (inverse) powers of $SO(5)$ breaking spurions, such that a light Higgs
does not necessarily lead to light top partners, resolving tensions with LHC searches.}

We will also discuss predictions of the setup for less clean observables in $ b \to s \ell^+ \ell^-$ decays
and take into account further flavor constrains on the model. We will in particular 
focus on the angular analysis of the $B \to K^\ast \mu^+ \mu^-$ decay, 
where pseudo-observables have been defined that also allow to cancel leading hadronic uncertainties \cite{Matias:2012xw, Jager:2012uw, Descotes-Genon:2013vna, Lyon:2014hpa, Descotes-Genon:2014uoa, Jager:2014rwa, Ciuchini:2015qxb, Capdevila:2017ert, Chobanova:2017ghn},
and where results  are available from {\it all} 3 LHC $pp$ experiments as well as from Belle, 
which are again pointing to a $\sim 4\,\sigma$
deviation from the SM. 
It turns out that non-negligible effects are also predicted in this decay,
allowing a significant improvement with respect to the SM, while still addressing the $R_{K^{(\ast)}}$ anomalies
and meeting the most stringent flavor bounds.

The remainder of this article is organized as follows.
In the next section, we will provide an analysis of the NP required to address the anomalies in LFU violating
decays in terms of $D=6$ operators, parametrizing heavy physics beyond the SM in a model-independent way.
We will then examine the structure of operators generated by the composite lepton model 
and present numerical predictions for $R_{K^\ast}$, scrutinizing the correlation with $R_K$ as well as $B_s - \bar B_s$
mixing and taking into account constraints from $B_s \to \mu^+\mu^-$. Finally,
we will give our predictions for the angular observable $P_5^\prime$, which was also found to feature a 
pronounced deviation from the SM prediction, 
before ending with our conclusions.

\section{Pattern of the $R_K, R_{K^\ast}$ Anomalies}

Employing the framework of effective field theory, a clear picture of the required structure of heavy
NP to explain the $R_{K^{(\ast)}}$ anomalies can be obtained. The relevant operators ${\cal O}_i$,
contributing to semi-leptonic $B_s$ decays to leading approximation, are contained in the effective Hamiltonian \cite{Grinstein:1988me, Buchalla:1995vs} 
\begin{equation}
\label{eq:Ham}
	{\cal H}_{\rm eff}^{\Delta B=1} = -\frac{4 G_F}{\sqrt 2} V_{tb}V_{ts}^\ast \frac{\alpha}{4 \pi} \!
	\sum_{\ i;\ell=e,\mu,\tau}\!\! \left( C_i^\ell {\cal O}_i^\ell + {C_i^\ell}^\prime {{\cal O}_i^\ell}^\prime\, \right) +{\rm h.c.}\,,
\end{equation}
and read 
\begin{equation}
\label{eq:Ops}
	{\cal O}_7=\frac{m_b}{e} (\bar s \sigma_{\mu\nu} P_R b) F^{\mu\nu} \,,\quad {\cal O}_9^\ell=(\bar s \gamma_\mu P_L b) (\bar 		\ell \gamma^\mu \ell) 
	\,,\quad {\cal O}_{10}^\ell=(\bar s \gamma_\mu P_L b) (\bar \ell \gamma^\mu \gamma_5 \ell)\,,
\end{equation}
and ${{\cal O}^\ell_{7,9,10}}^{\hspace{-6mm}\prime}\hspace{5mm}$ equivalently with $P_L \leftrightarrow P_R$, 
where $P_{L,R} \equiv (1\mp\gamma_5)/2$.\footnote{The scalar and pseudo-scalar operators 
${{\cal O}^\ell_{S,P}}^{\hspace{-3.5mm}(\prime)}$ are already considerably constrained from $B_s \to \mu^+ \mu^-$ and play no important role in the following discussions \cite{Hiller:2014yaa,Beaujean:2015gba}. This is also true for tensor operators,
which can not be generated from operators invariant under the SM gauge group to leading order \cite{Alonso:2014csa}.}
In fact, as the relevant energy scale is much below $m_W$, also the SM contributions are best expressed in
terms of contributions to the operators (\ref{eq:Ops}), $C=C^{\rm SM}+C^{\rm NP}$ , where (at $\mu=4.8$\,GeV) \cite{Capdevila:2016fhq}
\begin{equation}
	C_9^{\ell\,{\rm SM}}=4.07 \,, \ C_{10}^{\ell\,{\rm SM}}=-4.31 \,, \ C_7^{\rm SM}= -0.29\,,
\end{equation}
respecting LFU, {\it i.e.,} $C_i^{e\,{\rm SM}}=C_i^{\mu\,{\rm SM}}=C_i^{\tau\,{\rm SM}}$,
and ${{\cal O}^\ell_{7,9,10}}^{\hspace{-6mm}\prime\,{\rm SM}}=0$.

The operators above are written in terms of leptonic vector (${{\cal O}_9^\ell}^{(\prime)}$) and 
axial-vector (${{\cal O}_{10}^\ell}^{\!\!\!(\prime)}$) currents, which is convenient to add
higher oder corrections including SM gauge bosons (the photon couples vectorial).
NP, on the other hand, is conveniently parametrized in the chiral basis, writing the Hamiltonian (\ref{eq:Ham}) as 
\begin{equation}
	{\cal H}_{\rm eff}^{\Delta B=1} = -\frac{4 G_F}{\sqrt 2} V_{tb}V_{ts}^\ast \frac{\alpha}{4 \pi} \!
	\left( {\cal O}_7+{\cal O}_7^\prime +  \sum_{X,Y=L,R \atop \ell=e,\mu,\tau}\!\! \CXY{X}{Y}{\ell} \OXY{X}{Y}{\ell} \right) +{\rm h.c.}\,,
\end{equation}
with
\begin{equation}
	\OXY{X}{Y}{\ell} = (\bar s \gamma_\mu P_X b) (\bar \ell \gamma^\mu P_Y \ell)\,,
\end{equation} 
since often NP models treat a certain chirality in a special way and
one can take advantage of the accidental hierarchy of SM contributions 
\begin{equation}
	\CXYSM{L}{L}=8.38 \gg -\CXYSM{L}{R}=0.24
\end{equation}
to directly see the importance of NP effects via their interference with the SM contributions (see \cite{Hiller:2014ula, DAmico:2017mtc}). The coefficients of the two bases are related in a simple way via 
$C_9^\ell\!=\!(\CXY{L}{R}{\ell}\!+\!\CXY{L}{L}{\ell})/2$, $C_{10}^\ell\!=\!(\CXY{L}{R}{\ell}\!-\!\CXY{L}{L}{\ell})/2$, and similarly for the primed 
operators with $b\hspace{-0.12mm}s_{\hspace{-0.29mm}L} \to b\hspace{-0.12mm}s_{\hspace{-0.29mm}R}$. In this basis, we arrive at\footnote{The individual branching fractions are given by \cite{Ali:1999mm,Altmannshofer:2014rta,Hiller:2003js,Hiller:2014ula}
\begin{equation}
	\frac{d {\cal B}(B \to K \ell^+\ell^-)}{dq^2}=\tau_{B^+} \frac{G_F^2 \alpha^2 |V_{tb}V_{ts}^\ast|^2}{(4 \pi)^5 m_B^3}
	([m_{B\!-\!K^\ast}^2]^2\!-\!2m_{B\!+\!K^\ast}^2q^2\!+\!q^4)^{\frac 32}\,(|F_V|^2+|F_A|^2) \,,
\end{equation}
where
\begin{equation}
\begin{split}
	F_V(q^2)=&(\CXY{R}{R}{\ell}\!+\!\CXY{L}{L}{\ell}\!+\!\CXY{R}{L}{\ell}\!+\!\CXY{L}{R}{\ell})/2\, f_+(q^2) + \frac{2m_b}{m_B+m_K}
	(C_7 + C_7^\prime) f_T(q^2) + h_K(q^2)\,,\\
	F_A(q^2)=&(\CXY{R}{R}{\ell}\!-\!\CXY{L}{L}{\ell}\!-\!\CXY{R}{L}{\ell}\!+\!\CXY{L}{R}{\ell})/2\, f_+(q^2)\,,
\end{split}
\end{equation}
$m_{A\pm B}^2 \equiv m_A^2\pm m_B^2$, and we neglected lepton masses, CP violation,
and higher order corrections (which are however included in our numerical analysis, employing
CP averaged quantities).
Here, $f_+$ and $f_T$ are the QCD form factors (see \cite{Bouchard:2013pna,Buras:2014fpa}) 
and $h_K(q^2)$ parametrizes non-factorizable contributions from 
the weak hamiltonian \cite{Altmannshofer:2014rta}. Neglecting the strongly suppressed $C_7^{(\prime)}$ contributions
(which could only become relevant approaching the photon pole at $q^2 = 0$ and are in any case constrained
to be pretty SM like \cite{Paul:2016urs}) and the non-factorizable $h_K(q^2)$, the QCD form factor 
$f_+(q^2)$ drops out in the ratio $R_K$ (\ref{eq:RKEFT}).}
\begin{equation}
\label{eq:RKEFT}
	R_K=\frac{|\CXY{L}{L}{\mu}\!+\!\CXY{R}{L}{\mu}\!|^2+|\CXY{L}{R}{\mu}\!
	+\!\CXY{R}{R}{\mu}|^2}{|\CXY{L}{L}{e}\!+\!\CXY{R}{L}{e}\!|^2+|\CXY{L}{R}{e}\!+\!\CXY{R}{R}{e}|^2}\,.
\end{equation}

Given that we are considering corrections to $R_K$ of (up to) $\sim\!30\%$, the corresponding 
NP contributions to the Wilson coefficients will also be of this order, if they interfere
with the leading SM contribution $\CXYSM{L}{L}$ (barring significant cancellations). In consequence, 
it makes sense to expand $R_K$  to leading order
in the NP coefficients $\CXYNP{X}{Y}{\ell}$, to allow for a transparent theoretical interpretation, which we will do
below.
We keep however terms containing right handed lepton currents up to
quadratic order, going beyond the chiral linear approximation \cite{DAmico:2017mtc}, since they do not interfere with the 
leading (left-handed) SM contribution (for vanishing lepton masses). 
Thus, on the one hand, potentially larger effects are required to explain the anomalies (as happens in the explicit model
under consideration), and generically, due to the suppressed interference, quadratic terms become important.
Note however that (as long as $\CXYNP{X}{R}{\ell} < \CXYSM{L}{L}$) higher terms are still suppressed.\footnote{If quadratic terms in $\CXY{X}{R}{\ell}$
are kept, a $\sim 30\%$ effect in $R_K$ is {\it per se} in agreement with a convergence of the expansion 
in NP contributions, no matter from which operator it is induced
(see also \cite{Contino:2016jqw}).}
Neglecting also the strongly suppressed interference with $\CXYSM{L}{R}$, we arrive at 
\begin{equation}
\label{eq:expRK}
	\begin{split}
	R_K \simeq 1 + 2 & \frac{{\rm Re} [{\CXYSM{L}{L}}^{\!\!\!\!\ast} \, (\CXYNP{L}{L}{\mu}\!\!+\!\CXYNP{R}{L}{\mu}\!-\!\CXYNP{L}{L}		{e}\!\!-\!\CXYNP{R}{L}{e})]}{|\CXYSM{L}{L}|^2} \\[2.3mm]
	+&\textrm{\footnotesize$\frac{|\CXYNP{L}{R}{\mu}+\CXYNP{R}{R}{\mu}|^2-|\CXYNP{L}{R}{e}+\CXYNP{R}{R}{e}|^2}{|\CXYSM{L}{L}|^2}$}\,.
\end{split}
\end{equation}
For generic $30\%$ corrections to $R_K$, this formula is exact up to $\lesssim10\%$ corrections, 
which are smaller than the experimental uncertainty.\footnote{If 
the fourth order in $\CXY{X}{R}{\ell}$ is included, which corresponds to adding
\begin{equation}
	\Delta R_K^{(4)} = \frac{|\CXYNP{L}{R}{e}+\CXYNP{R}{R}{e}|^4-|\CXYNP{L}{R}{e}+\CXYNP{R}{R}{e}|^2
	|\CXYNP{L}{R}{\mu}+\CXYNP{R}{R}{\mu}|^2}{|\CXYSM{L}{L}|^4}\,,
\end{equation}
it holds even at the ${\cal O}(1\%)$ level, which becomes negligible compared to other uncertainties.
Still, for the numerical results presented in Section \ref{sec:CH}, we will use
the exact expressions, including in addition higher order QCD corrections \cite{david_straub_2017_569011} as well as the effect of $C_7^{\rm SM}$.}
In particular, it captures the {\it leading} effects of {\it all} NP Wilson coefficients considered.

We directly observe that $R_K<1$ can be realized in two ways. It could origin from a destructive (constructive)
interference of the combined 
left-handed muon (electron) contributions with the leading SM piece or, more generally, a negative sign
in the difference of muon and electron contributions $\CXYNP{L+R}{L}{(\mu-e)} \equiv 
\CXYNP{L}{L}{\mu}\!\!+\!\CXYNP{R}{L}{\mu}\!-\!\CXYNP{L}{L}{e}\!\!-\!\CXYNP{R}{L}{e}$.
On the other hand, it could stem from couplings to right-handed lepton currents. In that case, as discussed,
the quadratic NP contribution dominates in general. From (\ref{eq:expRK}) it then follows directly
that $R_K<1$ requires the effect to come from the electron sector.
Of course, a combination is possible, such that right-handed muon currents are allowed,
however, while for the case of electron currents, {\it any} operator alone could accommodate $R_K<1$,
for the muon case, right handed contributions {\it alone} are not feasible, no matter
what is the quark chirality.

More valuable information on the physical origin of the 
possible $B$-physics anomalies can be obtained by considering in addition the 
ratio $R_{K^\ast}$, which tests different combinations of Wilson coefficients, to which we will turn now.
The theoretical prediction in this case reads\footnote{Neglecting terms suppressed by $m_\ell^2/q^2$, NLO, and non-factorizable corrections (as well as CP violation),
the individual branching fractions can be expressed in terms of six transversity amplitudes $A_{0,\perp,\parallel}^{L,R}$
as \cite{Bobeth:2008ij} (see also \cite{Ali:1999mm,Hambrock:2013zya,Hiller:2014ula,Straub:2015ica}) 
\begin{equation}
	\begin{split}
	\frac{d {\cal B}(B^0 \to K^\ast \ell^+\ell^-)}{dq^2}\simeq & \tau_{B^0}  \frac{G_F^2 \alpha^2 |V_{tb}V_{ts}^\ast|^2}{3 \cdot (4 		\pi)^5 m_B^3}([m_{B\!-\!K^\ast}^2]^2\!-\!2m_{B\!+\!K^\ast}^2q^2\!+\!q^4)^{\frac 1 2}\, q^2\\
	& (|A_\perp^{\ell L}|^2+|A_\parallel^L|^2+|A_0^L|^2  + L\!\to\!R) \,,\\
\end{split}
\end{equation}
\vspace{0.4mm}
\begin{equation}
	\begin{split}
	A_\perp^{\ell L,R} & =+ \sqrt2 m_B (1\!-\!q^2/m_B^2) \left[ \CXY{L}{L,R}{\ell} + \CXY{R}{L,R}{\ell}  \right] \xi_\perp\,, \\
	A_\parallel^{\ell L,R} & =- \sqrt2 m_B (1\!-\!q^2/m_B^2)\left[ \CXY{L}{L,R}{\ell} - \CXY{R}{L,R}{\ell}  \right] \xi_\perp\,, \\
	A_0^{\ell L,R} & =- \frac{m_B^3 (1\!-\!q^2/m_B^2)^2}{2|q| m_{K^\ast}}\left[ \CXY{L}{L,R}{\ell} - \CXY{R}{L,R}{\ell}  \right] \xi_		\parallel\,,\\
	\end{split}
\end{equation}
where the form factors $\xi_{\perp,\parallel}$ are given, {\it e.g.}, in Appendix E of \cite{Bobeth:2008ij}.
We directly dropped electromagnetic dipole contributions, 
becoming important only for $q^2 \to 0$, which would appear in the three square brackets above as\textrm{
\scriptsize$2 m_b m_B/q^2 \{C_7+C_7^\prime,\,C_7-C_7^\prime,\,q^2/m_B^2 (C_7-C_7^\prime)\}$}. 

Defining the integrated form factors 
$$g_{\perp,\parallel,0}^{[q_{\rm min}^2,q_{\rm max}^2]}\!=
\!\int_{q_{\rm min}^2}^{q_{\rm max}^2}\!\!\!\! dq^2
\textrm{\scriptsize $([m_{B\!-\!K^\ast}^2]^2\!-\!2m_{B\!+\!K^\ast}^2q^2
\!+\!q^4)^{\frac 1 2} \frac{2(q^3 - m_B^2 q)^2}{m_B^2}$}\ \{ |\xi_\perp|^2,\,|\xi_\perp|^2,\,
\textrm{\scriptsize$\frac{(m_B^2\!-\!q^2)^2}{8q^2 m_{K^\ast}^2}$}|\xi_\parallel|^2 \}\,,$$ we can write $R_{K^\ast}$
as two combinations of Wilson coefficients, weighted by the polarization fraction 
$p \equiv \frac{g_0+g_\parallel}{g_0+g_\parallel+g_\perp}$, where $p \approx 0.86$ to good approximation for the $q^2$ range considered here (with some per cent deviation for the low $q^2$ bin) \cite{Hiller:2014ula,Bobeth:2008ij}.
}
\begin{equation}
\begin{split}
\label{eq:RKsEFT}
	\textrm{\small $R_{K^\ast}\!=\frac{(1\!-\!p)(|\CXY{L}{L}{\mu}\!\!+\!\CXY{R}{L}{\mu}|^2\!
	+\!|\CXY{L}{R}{\mu}\!\!+\!\CXY{R}{R}{\mu}|^2)
	+ p(|\CXY{L}{L}{\mu}\!\!-\!\CXY{R}{L}{\mu}|^2\!
	+\!|\CXY{L}{R}{\mu}\!\!-\!\CXY{R}{R}{\mu}|^2)}
	{(1\!-\!p)(|\CXY{L}{L}{e}\!\!+\!\CXY{R}{L}{e}|^2\!
	+\!|\CXY{L}{R}{e}\!\!+\!\CXY{R}{R}{e}|^2)
	+ p(|\CXY{L}{L}{e}\!\!-\!\CXY{R}{L}{e}|^2\!
	+\!|\CXY{L}{R}{e}\!\!-\!\CXY{R}{R}{e}|^2)}$}\,.\\
\end{split}
\end{equation}

Expanding again in $\CXYNP{X}{Y}{\ell}$, keeping quadratic terms only in $\CXYNP{X}{R}{\ell}$ and neglecting
the strongly suppressed interference with $\CXYSM{L}{R}$, we obtain
\begin{equation}
\begin{split}
	R_{K^\ast}\simeq R_K\, - &4p\frac{{\rm Re}[{\CXYSM{L}{L}}^{\!\!\!\!\ast}(\CXYNP{R}{L}{\mu}-\CXYNP{R}{L}{e})]}{|\CXYSM{L}{L}|^2}\\[2.3mm]
	- &\textrm{\footnotesize $4p\frac{{\rm Re}[\CXYNP{L}{R}{\mu}{\CXYNP{R}{R}{\mu}}^{\!\!\!\!\ast}-\CXYNP{L}{R}{e}
	{\CXYNP{R}{R}{e}}^{\!\!\!\!\ast}]}{|\CXYSM{L}{L}|^2}$}\,.
\end{split}
\end{equation}
From this expression it is evident that $R_{K^\ast}$ probes right-handed quark currents
since in their absence it becomes equivalent to $R_K$ to leading approximation. A similar conclusion holds in the absence of 
left-handed lepton currents, unless {\it both} left- and right-handed quark currents appear in 
LFU violating NP contributions.

\begin{figure}[!t]
	\begin{center}
	\includegraphics[height=2.8in]{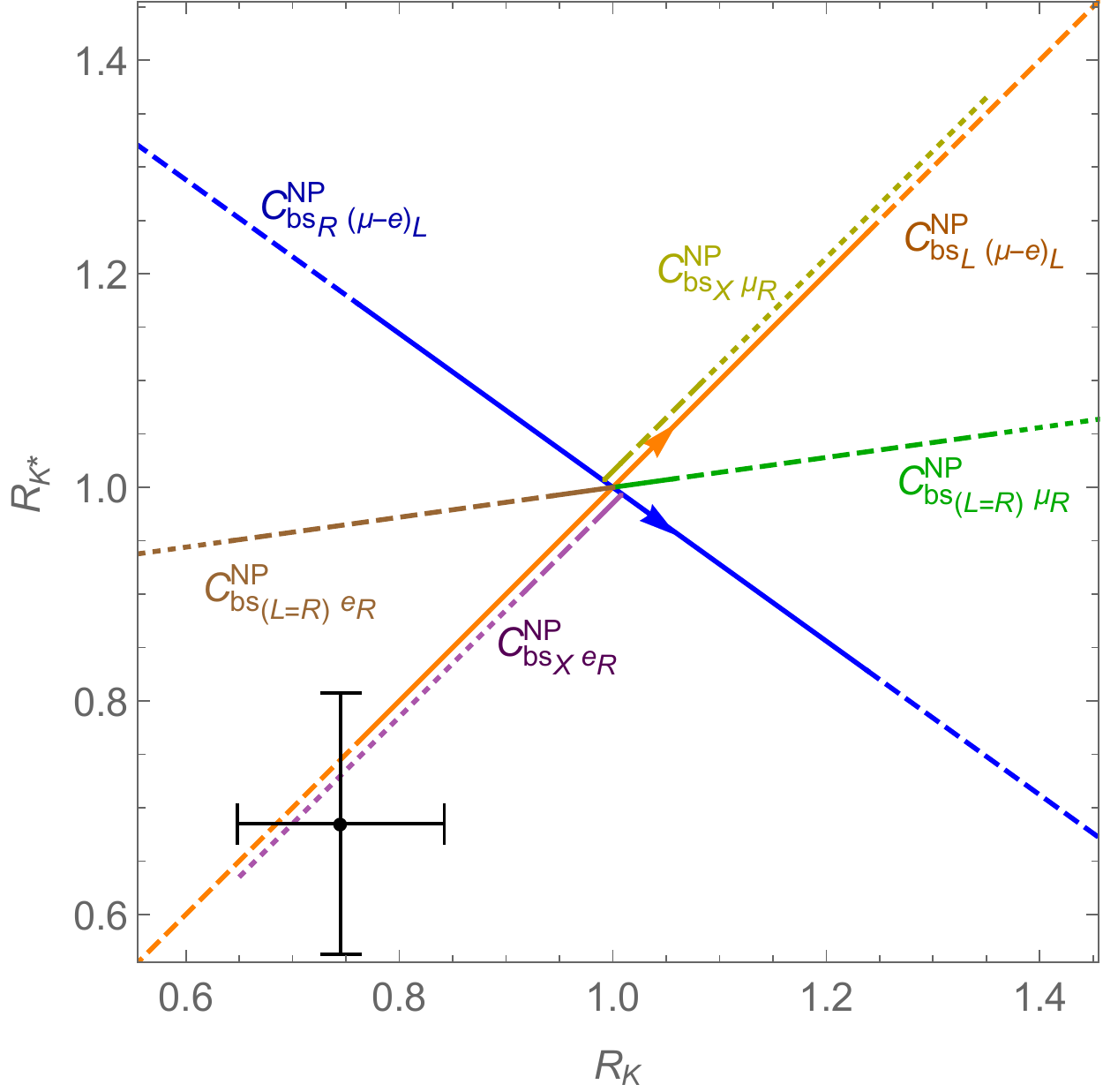} \quad \includegraphics[height=2.8in]{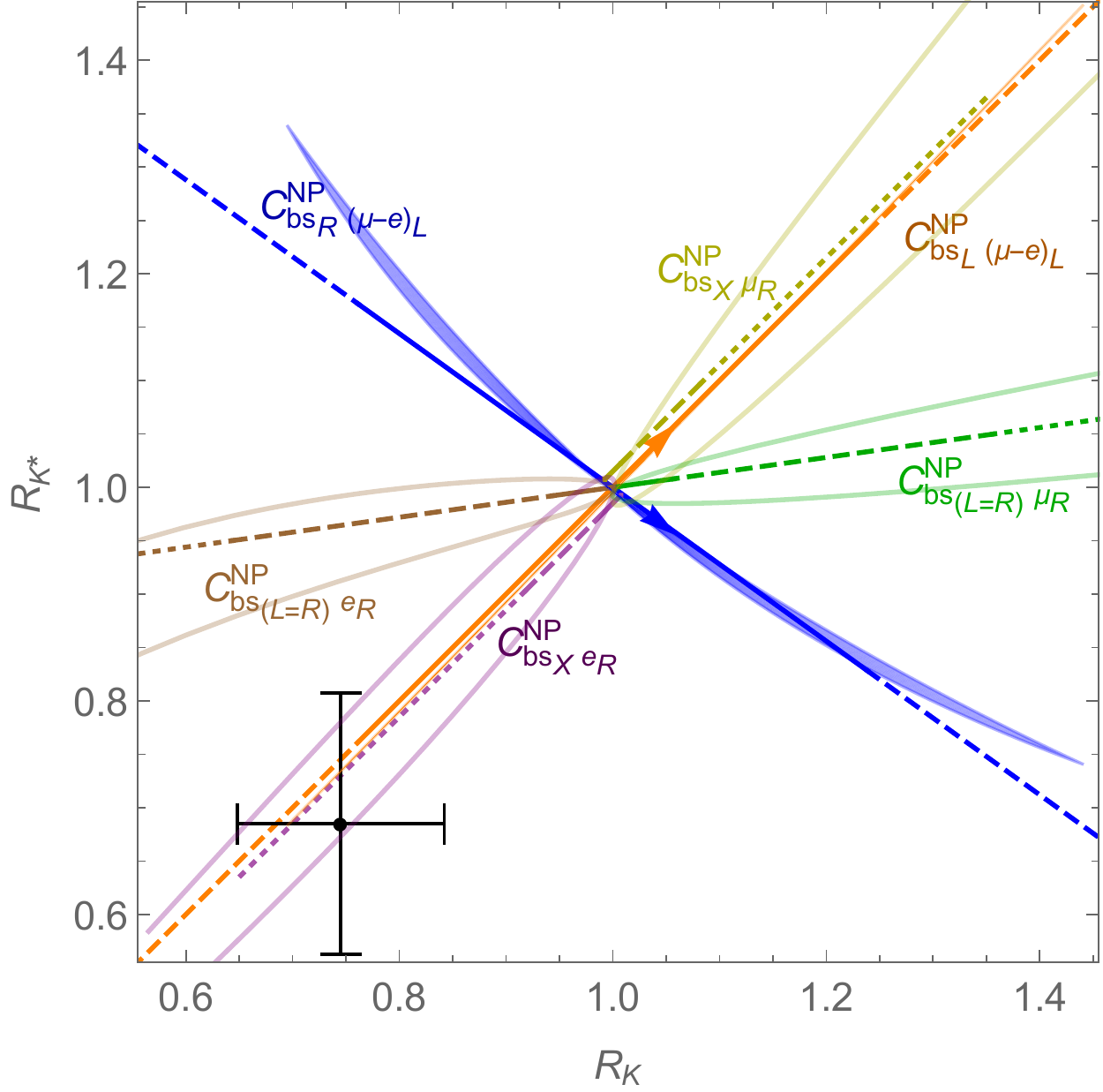} 
	\caption{\label{fig:RKs1} Predictions for $R_K$ vs. $R_{K^\ast}$ in dependence on the values
of the NP Wilson coefficients entering Eqs. (\ref{eq:RKEFT}) and (\ref{eq:RKsEFT}), where 
solid, dashed, and dotted lines correspond to $0 < |\CXYNP{X}{Y}{\ell}| < 1$,
$1 < |\CXYNP{X}{Y}{\ell}| < 2.5$, and $2.5 < |\CXYNP{X}{Y}{\ell}| < 5$, respectively. 
The $1\,\sigma$ allowed range from LHCb is depicted by the black uncertainty bars. 
In the right panel, the exact predictions 
are added as faint colored lines. See text for details.}
	\end{center}
\end{figure}

The findings above are visualized in Figure \ref{fig:RKs1}, where we show the correlations
between $R_K$ and $R_{K^\ast}$, employing Eqs. (\ref{eq:RKEFT}) and (\ref{eq:RKsEFT}). 
The colored lines correspond to the effects of the various NP Wilson coefficients.
We consider all the coefficients entering these expressions, including combinations, such as to 
allow simultaneously for NP in the muon and electron sectors and in different chirality combinations.
The most important dependencies of $R_K$ and $R_{K^\ast}$ are on the {\it difference}
of purely left-handed contributions involving muons and electrons $\CXYNP{L}{L}{(\mu-e)} \equiv 
\CXYNP{L}{L}{\mu} - \CXYNP{L}{L}{e}$, on the corresponding quantity involving right-handed quark
currents $\CXYNP{R}{L}{(\mu-e)} \equiv \CXYNP{R}{L}{\mu} - \CXYNP{R}{L}{e}$, where
the direction of positive values is indicated by an arrow,
and on the four coefficients involving right handed lepton currents, entering at quadratic order in NP.
Note that for the latter case, a {\it simultaneous} presence of left- and right-handed quark currents is necessary,
in order to break the degeneracy $R_K=R_{K^\ast}$, while in case only one coefficient is turned on, 
the effect of either of them is indistinguishable in the $R_K$ vs. $R_{K^\ast}$ plane.
We thus consider the distinct individual contributions $\CXYNP{X}{R}{\mu}$ and $\CXYNP{X}{R}{e}$
(being equal for $X=L,R$)
as well as the simultaneous presence of left- and right-handed quark currents
via $\CXYNP{L}{R}{\ell}=\CXYNP{R}{R}{\ell}\equiv \CXYNP{(L=R)}{R}{\ell}$,
to capture the most relevant different scenarios.
The effect of further combining different contributions to $R_K$ and $R_{K^\ast}$ can be easily inferred by
considering the analytic Eqs. (\ref{eq:RKEFT}) and (\ref{eq:RKsEFT}) in addition to the figure.
The size of the coefficients corresponding to a certain point in the plane is visualized via the shape of the lines
-- solid lines correspond to $0 < |\CXYNP{X}{Y}{\ell}| < 1$, dashed lines to $1 < |\CXYNP{X}{Y}{\ell}| < 2.5$,
while dotted lines feature $2.5 < |\CXYNP{X}{Y}{\ell}| < 5$. Since the interference of NP effects
in the right-handed lepton sector with SM contributions is suppressed, generically larger coefficients are required
here in order to obtain sizable effects.

From the plot it becomes clear that the LHCb results, visualized by the black uncertainty bars,
uniquely single out either a left-handed-left-handed NP effect, 
where $\CXYNP{L}{L}{\mu} (\CXYNP{L}{L}{e})$ needs to feature a negative (positive) 
sign, or an effect in the right-handed {\it electron} sector via $\CXYNP{X}{R}{e}$,
as the preferred solution to the $R_K^{(\ast)}$ anomalies.\footnote{It becomes also 
evident that a negative NP contribution to 
$C_9^\mu\!=\!(\CXY{L}{R}{\mu}\!+\!\CXY{L}{L}{\mu})/2$, as advocated as a solution 
to the $B \to K^\ast \mu^+ \mu^-$ anomaly (see, {\it e.g.,} \cite{Descotes-Genon:2013wba,Altmannshofer:2017fio}), allows for a good fit of the $R_K^{(\ast)}$
anomalies, basically because (for moderate values of the coefficients) the 
effect of $\CXYNP{L}{L}{\mu}$ dominates via the SM interference.
A positive $C_9^e$, on the other hand, also allows for a straightforward solution.}
This is a very interesting finding with respect to the model considered in the 
remainder of the paper. In fact, while a number of models accommodate
the former option of dominating left-handed effects (including leptonic vector currents) 
\cite{Crivellin:2015mga, Crivellin:2015lwa, Niehoff:2015bfa, Sierra:2015fma, Celis:2015ara, Greljo:2015mma, Falkowski:2015zwa, Allanach:2015gkd, Chiang:2016qov, Boucenna:2016wpr, Megias:2016bde,  Altmannshofer:2016jzy, Crivellin:2016ejn, GarciaGarcia:2016nvr, Bhatia:2017tgo, Bonilla:2017lsq, Alonso:2017uky, Ellis:2017nrp, Tang:2017gkz, Chiang:2017hlj, King:2017anf, Buttazzo:2017ixm, Megias:2017vdg, Cline:2017ihf}, the latter solution hardly 
exists in the form of explicit models in the literature, however just emerges in the composite model
presented in Section \ref{sec:CH}.

We finally stress again that for the predictions of the model discussed in the next section,
we employ full results including higher order corrections and the (small) effects of 
$C_7^{\rm SM}$, performing a quadratic fit in the NP Wilson coefficients to the NLO results.\footnote{We used the code \textsc{Flavio} (v0.21.1) \cite{david_straub_2017_569011} for the numerical
prediction.}
We also display, in the right panel of Figure \ref{fig:RKs1}, the results employing these
full expressions, visualized via faint colored lines. 
Note that now, in the case of right-left contributions,
the prediction does not just depend on the difference of muon and electron contributions
any more and varying $\CXYNP{R}{L}{\mu}$ and $\CXYNP{R}{L}{e}$ independently leads to a (very modest) 
spread of the predictions in the $R_K$ vs. $R_{K^\ast}$ plane, depicted by the blue shaded region.
Generally, the approximate results describe the relevant physics quite accurately.

In summary, consistently explaining the $R_K$ and $R_{K^\ast}$ anomalies
requires both quark FCNCs involving the $b$ quark
and LFU violation in the $\mu$ vs.~$e$ system 
(with effects either in left-handed currents in both sectors, or with
a non-negligible {\it right-handed electron} contribution).
The model that we will discuss now naturally leads
to both effects, via exchange of composite vector resonance,
whose couplings are not aligned with the SM couplings
and the biggest contributions are actually expected in quark transitions 
involving the third generation and LFU violation involving light SM leptons.
Since larger corrections are predicted for electrons, the setup
matches nicely with the fact that this sector is less constrained, 
and, as we will see, sizable effects are in fact possible without problems with,
{\it e.g.}, $B_s \to \ell^+ \ell^-$ decays.

\section{Predictions in Composite Framework and Further Observables}
\label{sec:CH}

Composite Higgs models offer \emph{a priori} a compelling framework to explain the neutral flavor anomalies. The presence of a rich spectrum of bound states at the $\TeV$ scale, including heavy vector resonances (of $Z^{\prime}$ type) with  sizable couplings to some of the SM fermions, make these scenarios natural candidates to address the tension between data
and SM predictions. Moreover, and contrary to most of the solutions to these anomalies that one can find in the literature, they also offer an interplay with  electroweak symmetry breaking (EWSB) and some rationale to solve the hierarchy problem. If one considers that fermion masses are generated via the mechanism of \emph{partial compositeness}, a sizable violation of lepton flavor universality, like hinted by $R_K$ and $R_{K^{\ast}}$, necessarily requires the charged leptons to feature a sizable degree of compositeness in their left-handed (LH) and/or right-handed (RH) chirality, $\epsilon_{\ell_R}$, $\epsilon_{\ell_L}$.  Since charged lepton masses scale in general as $ \sim g_{\ast} v\epsilon_{\ell_L} \epsilon_{\ell R}$, where $g_{\ast}$ is the characteristic coupling within the strong sector, both chiralities can not be composite at the same time. Therefore, all these models will either predict effects in $\mathcal{O}_{bs_{L,R} \ell_L}$ or in $\mathcal{O}_{bs_{L,R}\ell_R}$ scaling like $\sim g_{\ast}^2/m_{\ast}^2 V_{ts}\epsilon_{b_X}^2 \epsilon_{\ell_Y}^2$, where $X$ and $Y$ denote the possible chiralities involved
in the quark- and lepton-sector, respectively, and $m_{\ast}$ is the typical mass scale of the first vector resonances.  

The model under consideration falls into the second category and was presented in detail in \cite{Carmona:2015ena, Carmona:2014iwa}. One of its most interesting features is that charged leptons partially substitute the role of the top quark as a trigger of EWSB, and a link between violation of lepton flavor universality and the absence of top partners at the LHC is established. Indeed, if the composite operators interacting with the RH charged leptons transform in sufficiently large irreducible representations of the global group within the strong sector,  the leading charged lepton contribution to the Higgs quartic 
coupling will appear at order $\sim |\epsilon_{\ell_R}|^2$ instead of the usual $\sim |\epsilon_{\ell_R}|^4$. Since the leading top contribution can be expected to appear at order $|\epsilon_{t_L}|^4$, the contribution arising from the lepton sector can be comparable to the top one, even with a smaller degree of compositeness. Moreover, if all the three lepton generations are partially composite, the lepton contribution will be enhanced by a factor $N_{\rm gen}\sim  3$, compensating the color factor $N_c = 3$ present in the top case and allowing to lift the top partners via destructive interference.  One of the important findings of \cite{Carmona:2015ena, Carmona:2014iwa} is that the very same representations making this possible also provide the required quantum numbers for a minimal implementation of a type-III seesaw mechanism for the neutrino masses, which can
further motivate RH charged-lepton compositeness, as we will see now.\,\footnote{This is true at least for the $SO(5)/SO(4)$ and the $SO(7)/G_2$ \cite{Ballesteros:2017xeg} cosets.}  If one follows this very minimal avenue, considering each generation of RH leptons to interact with a {\it single} composite operator, all RH charged leptons in fact inherit the degree of compositeness of their RH neutrino counterparts, and the latter is required to be sizable to allow for large 
enough neutrino masses (see \cite{Carena:2009yt, delAguila:2010vg, Hagedorn:2011pw, Carmona:2015ena, Carmona:2014iwa, Carmona:2016mjr} for more details in both type-I and type-III seesaw models). Then, the different scaling of the neutrino and charged lepton mass matrices requires  $0\ll \epsilon_{\tau_R}\ll \epsilon_{\mu_R} \ll \epsilon_{e_R}$, in order to have simultaneously hierarchical charged lepton masses and a non-hierarchical neutrino mass matrix \cite{Carmona:2015ena}. An immediate consequence of the above chirality structure is that mostly the operators $\OXY{X}{R}{e}$ will be generated (as well as subdominantly $\OXY{X}{R}{\mu}$). Moreover, possible modifications of $Z$ couplings which are extremely constrained by electroweak precision data are avoided due to custodial symmetry, contrary to what happens for the case of composite LH leptons, where it is not possible to protect the coupling to both fields in the SM doublet at the same time.

To be concrete, we are considering a strongly interacting sector featuring the Higgs as a pseudo Nambu-Goldstone boson (pNGB) arising from the symmetry breaking $SO(5)\to SO(4)$, known as the Minimal Composite Higgs model (MCHM) \cite{Contino:2003ve, Agashe:2004rs}.
The quark fields
are embedded in ${\bf 5}^u_L,{\bf 1}^u_R,{\bf 5}^d_L,{\bf 1}^d_R$ representations of $SO(5)$,
while {\it all} lepton fields are embedded in only {\it two} representations, ${\bf 5}_L^\ell, {\bf 14}_R^\ell$, 
per generation. As mentioned before, we can explaine the tiny neutrino masses via a type-III seesaw mechanism, since the symmetric representation ${\bf 14}_R^\ell \cong \bf{(1,1) + (2,2) +  (3,3)}$  of $SO(5) \cong SU(2)_L \times SU(2)_R$ can host both an electrically charged $SU(2)_L$ singlet lepton ($\ell_R$) 
and a heavy fermionic seesaw triplet of $SU(2)_L$. This {\it unification} of right handed leptons
comes along with a more minimal quark representation than in known models, because the enhanced leptonic contribution 
to the Higgs potential, originating from the symmetric $SO(5)$ representation, allows for a viable electroweak 
symmetry breaking with all right handed SM-quarks inert under $SO(5)$ and the left-handed ones in the 
fundamental (see \cite{Carmona:2015ena,Carmona:2014iwa} for details)\footnote{The leptonic contribution also leads to a viable Higgs mass, 
without the need for ultra-light top partners \cite{Carmona:2014iwa} which would be problematic in the light
of LHC searches.}. Thus, the model features less degrees of freedom than standard incarnations, such
as the MCHM$_5$. 

One of the main challenges of all these scenarios featuring composite leptons is the generation of dangerous FCNCs through the exchange of the very same vector resonances producing the violation of LFU. Since, in general, they will also contribute to extremely well measured lepton flavor violating processes like $\mu\to e\gamma$, $\tau\to \mu \gamma$, $\mu \to eee$, and $\mu-e$ conversion, they typically require the addition of some non-trivial flavor symmetry. In the model at hand, it turns out that the reduced number of composite operators mixing with the light leptons naturally allows for a very strong flavor protection along the lines of  minimal flavor violation,  since a single spurion allows to break the flavor symmetry \cite{Carmona:2015ena}.
Regarding the quark sector, in the model at hand the left handed current $\bar s_L \gamma^\mu b_L$  ({\it i.e.,} $X=L$)
will in general dominate, with the compositeness of $b_L$ following from the large top mass, but also non-negligible contributions from right-handed quarks are possible. 

Let us conclude this discussion by stressing that if the effect of the model would be mostly due to the muon instead of the electron,
neither the $R_{K^\ast}$ nor the $R_K$ anomaly could be addressed. Once the effect originates from a right-handed
lepton current, it is required that the electron channel dominates in order to resolve the anomalies,
see Figure \ref{fig:RKs1}.
On the other hand, contrary to the case of NP mostly in the muon current, in this case 
both chiralities for the quark current are fine.
Thus, the very peculiar pattern of effects in our composite lepton scenario fits very well with experimental observation,
predicting 
\begin{equation}
\label{eq:RKRKs1}
R_K \sim R_{K^\ast} <1 \,.
\end{equation}

\begin{figure}[!t]
	\begin{center}
	\includegraphics[height=2.95in]{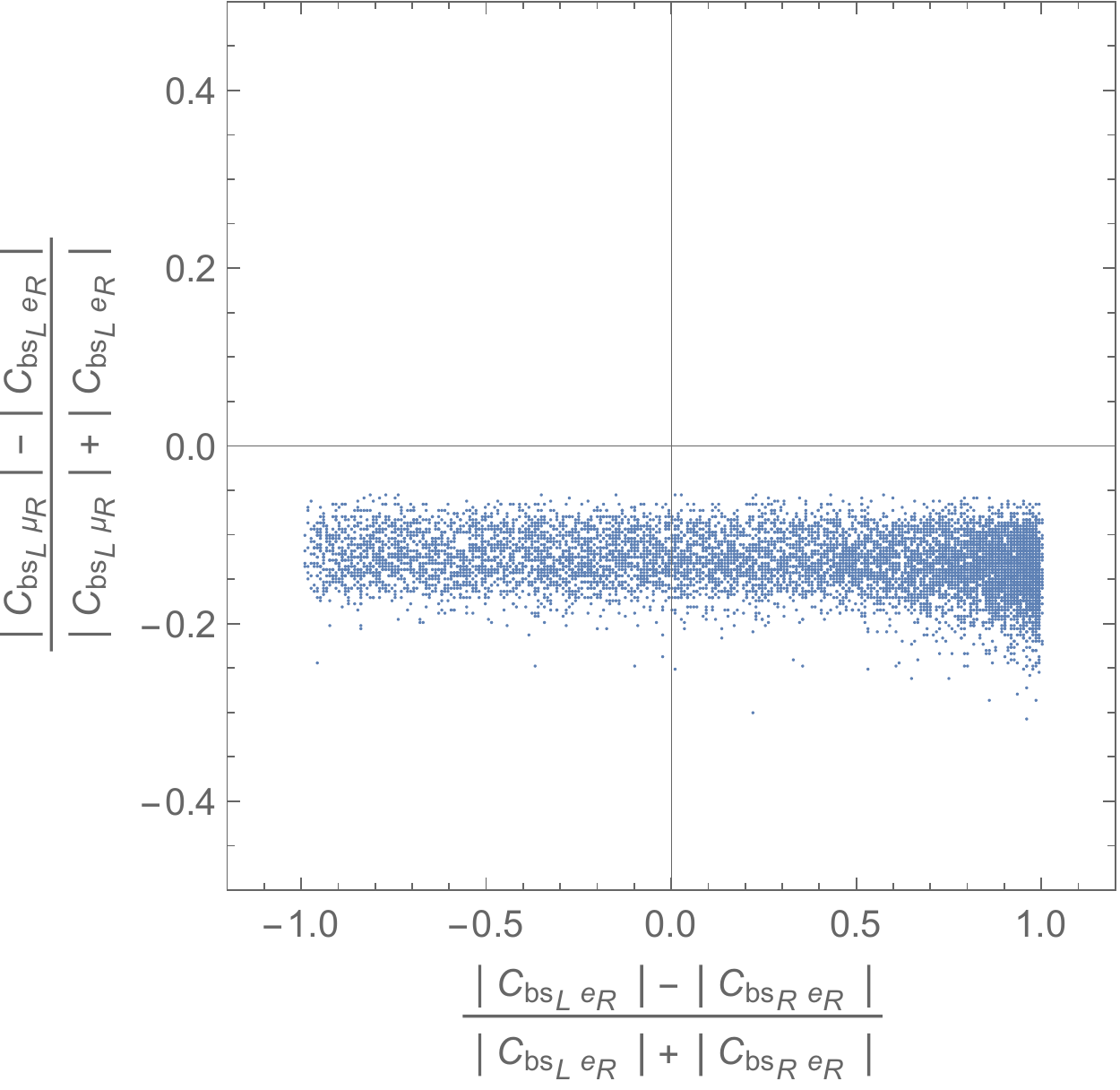}
	\caption{\label{fig:Wilsons}  Pattern of coefficients generated in the composite model. See text for details.}
	\end{center}
	\end{figure}

Before presenting our final numerical results for these observables, we show in Figure \ref{fig:Wilsons}
the pattern of Wilson coefficients generated in the composite Higgs model.
Here, and in the following, the points shown correspond to a scan over the parameter space of a 5D gauge-Higgs-unification
description of the composite Higgs framework detailed above. Brane and bulk masses have been generated randomly
(see \cite{Carmona:2015ena} for more details), requiring the correct SM spectrum to emerge at low energies, while the pNGB 
decay constant has been chosen as $f=1.2$\,TeV, such as to solve the hierarchy problem while being in agreement with electroweak precision data. The large density of points in the lower right quadrant confirms numerically the expectation lined out before of
$\mathcal{O}_{bs_{L} \ell_R}$ dominating in general, while in some regions of parameterspace
also  $\mathcal{O}_{bs_{R} \ell_R}$ can be significant, all leading to the pattern (\ref{eq:RKRKs1}).

\begin{figure}[!t]
	\begin{center}
	\includegraphics[height=2.75in]{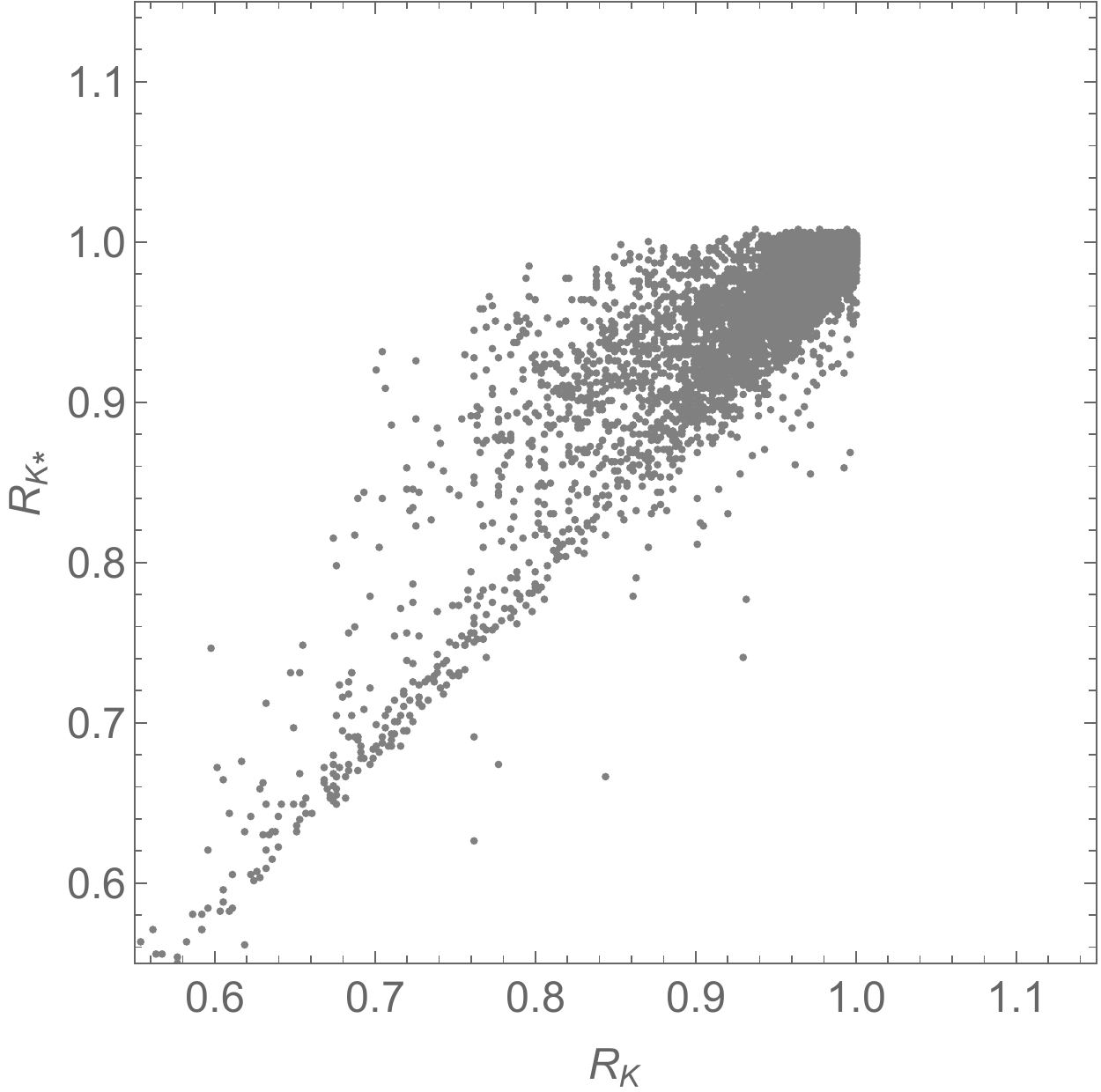} \quad  \includegraphics[height=2.75in]{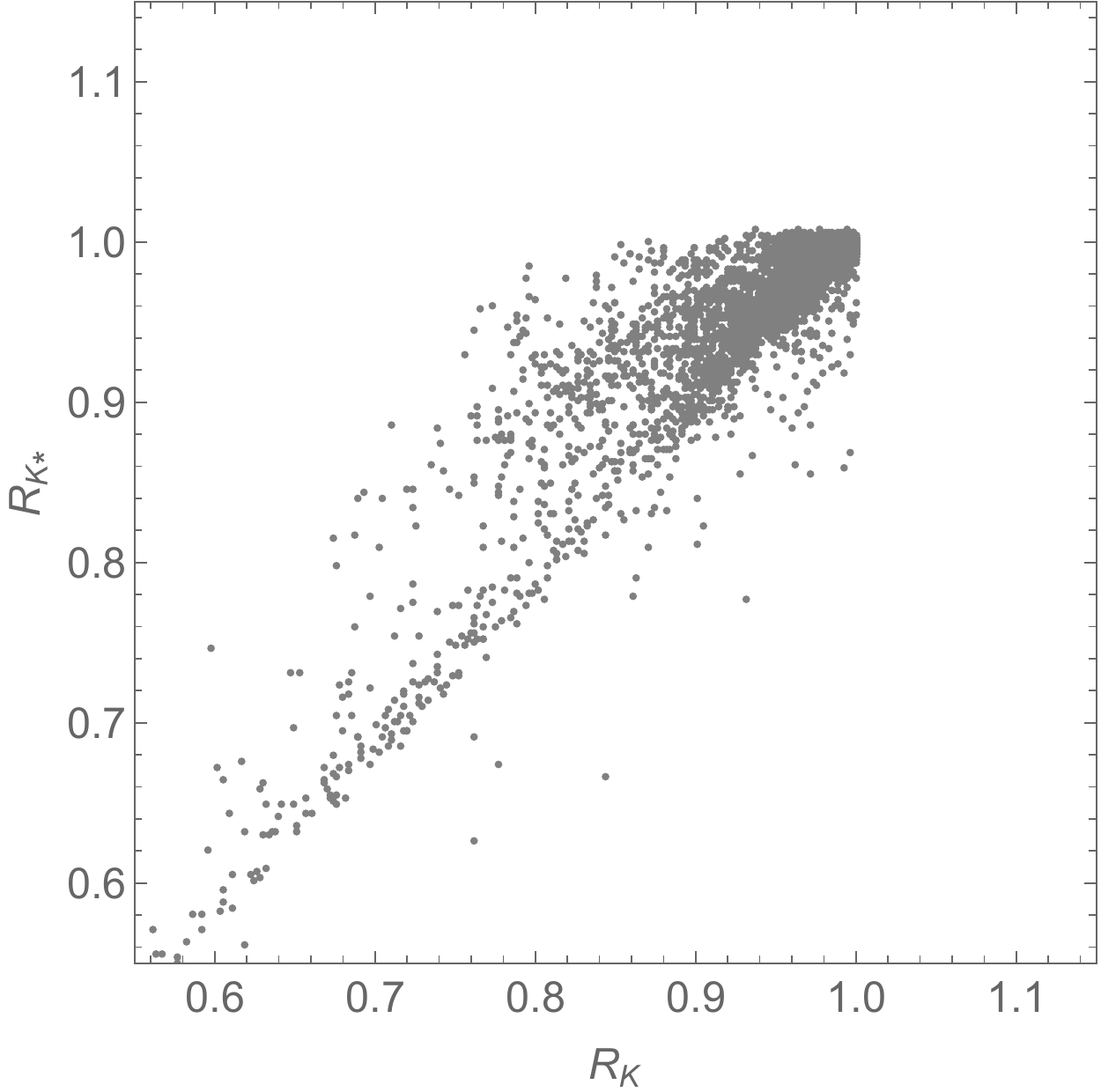}
	\caption{\label{fig:RKRKs} Predictions in the $R_K - R_{K^\ast}$ plane. In the right plot,
	points that feature a tension with the constraints from $B_s \to \mu^+\mu^-$ of more than 
	$1\,\sigma$ are rejected. See text for details.}
	\end{center}
	\end{figure}

After this general discussion, we present our predictions in the $R_{K^\ast}$ vs. $R_K$ plane in the left
panel of Figure \ref{fig:RKRKs}. One can clearly observe the sought pattern of $R_K \sim R_{K^\ast} <1$, emerging as discussed in detail above. 

Since along with $\mathcal{O}_{bs_{X} e_R}$ also, subleadingly, $\mathcal{O}_{bs_{X} \mu_R}$ is generated,
all the points need to face constraints from the rather well-measured branching ratio ${\cal B}(B_s \to \mu^+\mu^-)$,
which, in the chiral basis, is given by \cite{DeBruyn:2012wk}\footnote{Here, we neglect scalar and pseudoscalar operators, not generated in the model at hand to good approximation.}
\begin{equation}
\label{eq:Bsll}
	\frac{\!\!\!\!{\cal B}(B_s \to \ell^+\ell^-)}{\ {\cal B}(B_s \to \ell^+\ell^-)_{\rm SM}} = 
	\left|1+\frac{\CXYNP{L}{R}{\ell}\!-\!\CXYNP{L}{L}{\ell}
	\!-\!\CXYNP{R}{R}{\ell}\!+\!\CXYNP{R}{L}{\ell}}{\CXYSM{L}{R}\!-\!\CXYSM{L}{L}}\right|^2 \,,
\end{equation}
where \cite{Bobeth:2013uxa,Aaij:2017vad,Aaltonen:2009vr}
\begin{eqnarray}
	{\cal B}(B_s \to \mu^+ \mu^-)_{\rm SM} &=& (3.65\pm0.23) \times 10^{-9}\,,\\
	{\cal B}(B_s \to \mu^+ \mu^-)_{\rm exp} &=& (3.0\pm0.6^{+0.3}_{-0.2}) \times 10^{-9}\,,\\
	{\cal B}(B_s \to e^+ e^-)_{\rm SM} &=& (8.54 \pm 0.55) \times 10^{-14}\,,\\
	{\cal B}(B_s \to e^+ e^-)_{\rm exp} &<& 2.8 \times 10^{-7}\ @\,90\%\,{\rm CL}\,.
\end{eqnarray}
Note that in general the muonic final states are more stringently constrained experimentally, 
leaving more room for effects in the electron channel, which does not constrain the model at hand at all.
The former, on the other hand, has a (mild) impact on the setup. In general, avoiding effects in 
$B_s \to \ell^+ \ell^-$, requires either a leptonic or a quark vector-current - in fact the decay
tests products of axial-vector currents. In consequence, the negative $\CXYNP{L}{L}{\mu}$
solution to the $R_{K^{(\ast)}}$ anomalies could be accompanied by a $\CXYNP{R}{L}{\mu}$ contribution, 
to cancel effects in $B_s \to \ell^+ \ell^-$, which would however go in the wrong direction with respect
to $R_{K^\ast}$. The second option is an addition of $\CXYNP{L}{R}{\mu}$, which is less problematic since 
suppressed by the smaller interference with the SM, and corresponds to the $C_9$ solution of the
$b \to s \ell^+ \ell^-$ anomalies.
Note that to go into the direction of the very modest (negative) deviation from the SM prediction in $B_s\to \mu^+ \mu^-$, for a fixed chirality 
for one fermion current, the other should feature the larger effect in the opposite chirality 
(for positive Wilson coefficients), {\it i.e.}, $\CXYNP{X}{Y}{\mu}-\CXYNP{X}{X}{\mu}>0$
or $\CXYNP{Y}{X}{\mu}-\CXYNP{X}{X}{\mu}>0$, $(X,Y)\in\{(L,R),(R,L)\}$ (see eq. (\ref{eq:Bsll})).
This means in particular that an effect stemming solely from a negative $\CXYNP{L}{L}{\mu}$ 
goes in principle in a favorable direction.

	\begin{figure}[!t]
	\begin{center}
	\includegraphics[height=2in]{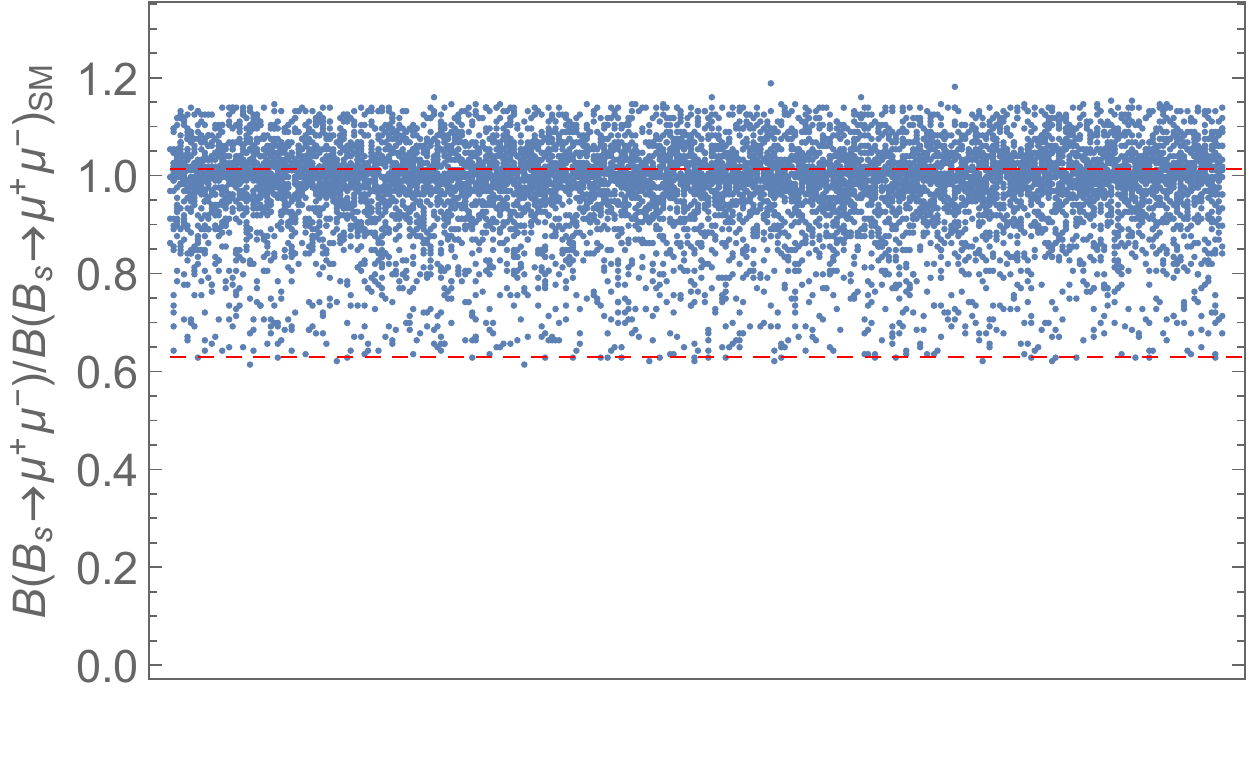} 
	\caption{\label{fig:Bsmumu} Predictions for ${\cal B}(B_s \to \mu^+ \mu^-)/{\cal B}(B_s \to \mu^+ \mu^-)_{\rm SM}$ in 
	the composite Higgs scenario. The experimental $1\,\sigma$ range is depicted by the red dashed lines.}
	\end{center}
	\end{figure}
For the composite Higgs model considered, effects are coming from non-vanishing $\CXYNP{L-R}{R}{\mu}, $
but are modest in general, as can be seen from Figure \ref{fig:Bsmumu} where we show the ratio
${\cal B}(B_s \to \mu^+ \mu^-)/{\cal B}(B_s \to \mu^+ \mu^-)_{\rm SM}$, together with the experimental
$1\,\sigma$ range, the latter depicted by the red dashed lines. Most of the points lie within the corresponding range
(while basically all meet the $2\,\sigma$ constraints). The impact of the $1\,\sigma$ bound on the
composite solution to the $R_{K^{(\ast)}}$ anomalies is visualized in the right panel of Figure \ref{fig:RKRKs},
where we reject the parameter-space points that do not meet this constraint. As can be seen, the effect
is very modest.

We now move to study the correlation with $B_s - \bar B_s$ mixing, focusing on the mass difference
between the heavy and light mass eigenstate $\Delta M_{B_s} \equiv M_H^s - M_L^s$ (see e.g. \cite{Buras:1998raa}).
The relevant additional operators, entering $B_s - \bar B_s$ mixing, are contained in the $\Delta B=2$ Lagrangian \cite{david_straub_2017_569011}
\begin{equation}
{\cal H}_{\rm eff}^{\Delta B=2} =\, -\!\!\!\!\! \sum_{XY\!=\!LL,LR,RR} \!\!\!\!\! C_V^{XY} {\cal O}_V^{XY} +{\rm h.c.}\,,
\end{equation}
where 
\begin{equation}
{\cal O}_V^{XY} = (\bar s \gamma_\mu P_X b) (\bar s \gamma^\mu P_Y b)\,,
\end{equation}
and are generated predominantly via the exchange of gluon resonances in the composite Higgs framework.
Note that we neglect again scalar (and tensor) operators, not present for the model at hand to good approximation. 
Our general prediction, obtained via a fit to the numerical results \cite{david_straub_2017_569011}
at the scale $\mu=2\,$TeV, reads 
\begin{equation}
	\frac{\Delta M_{B_s}}{\Delta M_{B_s}^{\rm SM}} \simeq 
	1 + \left( 35\,380\,{\rm Re}\,C_V^{LR\,{\rm NP}} - 10\,530\,{\rm Re}[ C_V^{LL\,{\rm NP}} + C_V^{RR\,{\rm NP}}] \right){\rm TeV}^2\,,
\end{equation}
where $\Delta M_{B_s}^{\rm SM}=1.313\times 10^{-11}\,$GeV.

\begin{figure}[!t]
	\begin{center}
	\includegraphics[height=2.75in]{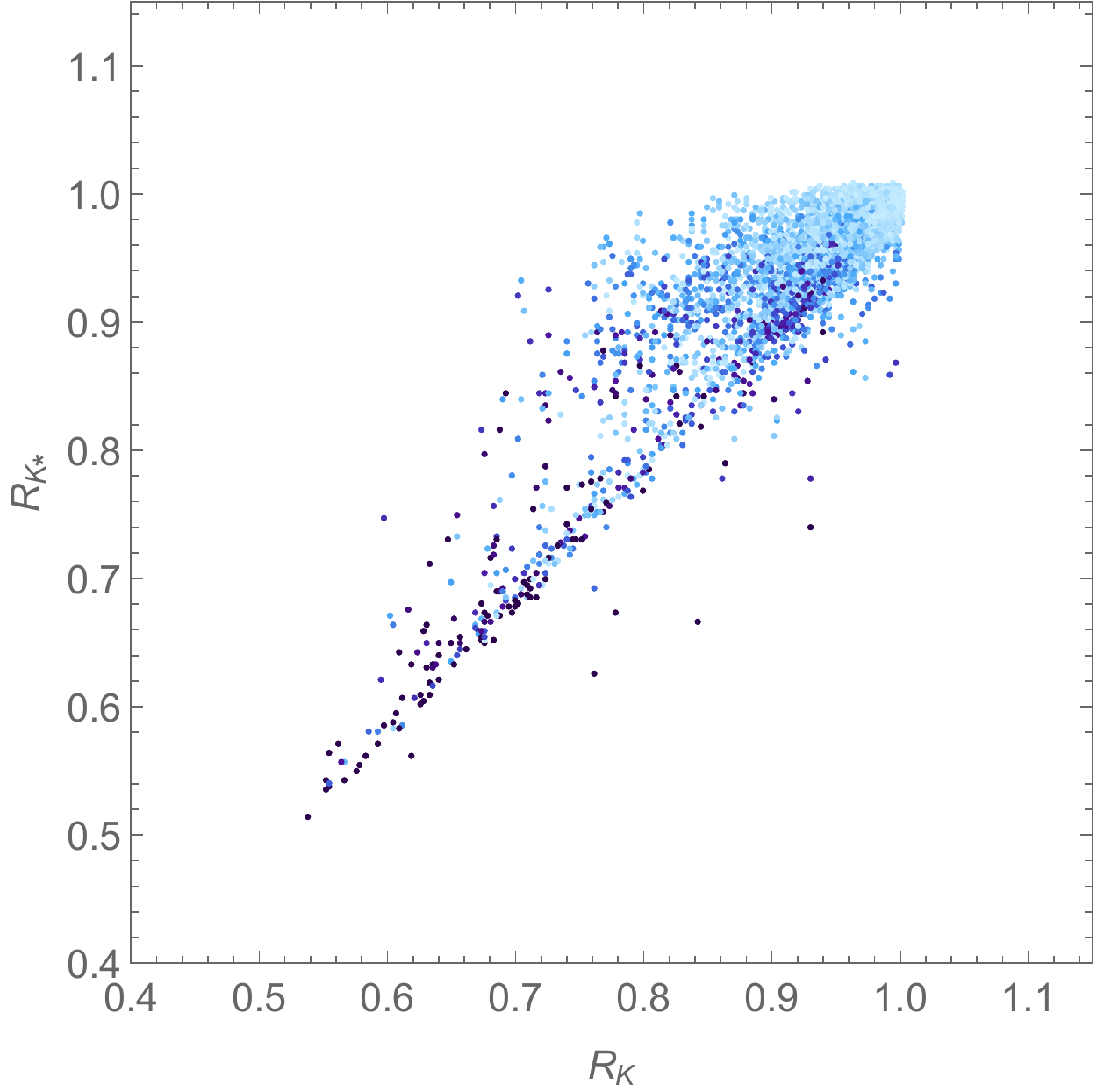} \raisebox{0.69cm}{\includegraphics[height=2.65in]{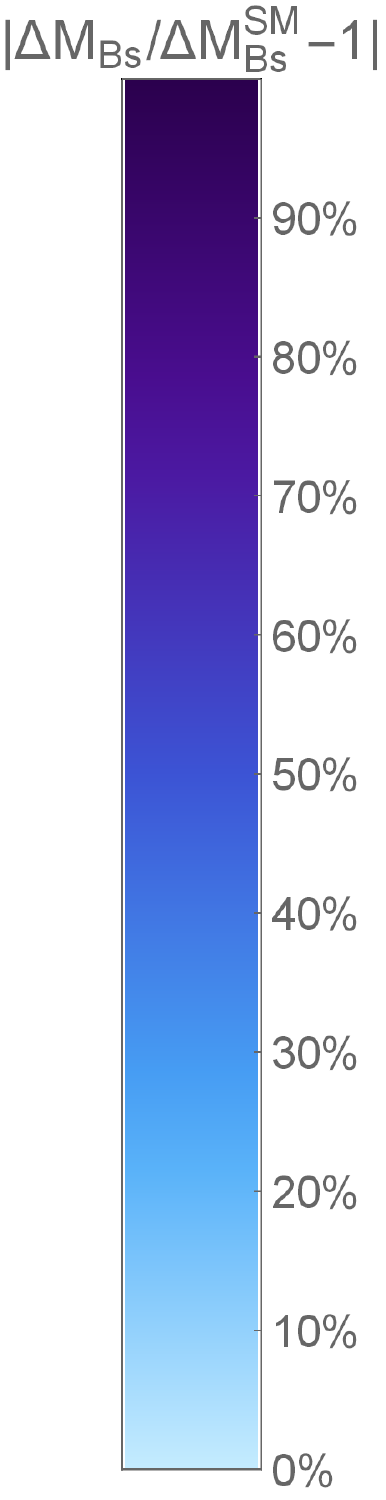}}
	\caption{\label{fig:MB} Predictions in the $R_K - R_{K^\ast}$ plane, where the colors depict the
	size of the corrections to  $\Delta M_{B_s}$. See text for details.}
	\end{center}
	\end{figure}

In Figure \ref{fig:MB} we provide again the results in the $R_{K^\ast}$ vs. $R_K$ plane, where now the
color code depicts the size of the corrections to the SM value of $\Delta M_{B_s}$, with light (dark) 
blue corresponding to modest (sizable) modifications. We find that there is a considerable amount 
of points with $(20-30)\%$ effects in $R_{K^{(\ast)}}$ and deviations in $\Delta M_{B_s}$ at the $\lesssim 20\%$
level, which is in the ballpark of the uncertainty in the SM prediction.
In consequence, values $R_K \sim R_{K^\ast} \sim (0.7-0.8)$ are possible, while constraints
from  $B_s - \bar B_s$ mixing are met. Since small deviations in $\Delta M_{B_s}$ can be achieved
via a cancellation between LH and RH quark currents, the presence of both $\CXYNP{L}{X}{\ell}$
and $\CXYNP{R}{X}{\ell}$ is somewhat preferred, moving away from the $R_K = R_{K^\ast}$
line (see Figure \ref{fig:RKs1}).\footnote{Note that solutions to 
the $R_K$ anomaly are also subject to competitive
constraints from searches for tails in the high $p_T$ di-lepton spectrum \cite{Greljo:2017vvb}. 
However, due to the suppressed couplings to light quarks, our setup avoids these bounds.}

Finally, another interesting set of observables in the field of semi-leptonic $B$
decays are the angular dependencies of the $B \to K^\ast \mu^+ \mu^-$
decay rate. Of particular interest is the coefficient $P_5^\prime$, which 
belongs to a class of observables constructed such as to cancel hadronic
uncertainties \cite{Matias:2012xw, Jager:2012uw, Descotes-Genon:2013vna, Lyon:2014hpa, Descotes-Genon:2014uoa, Jager:2014rwa, Ciuchini:2015qxb, Capdevila:2017ert, Chobanova:2017ghn}, which shows an interesting deviation with respect
to the SM prediction, see {\it e.g.}, \cite{Descotes-Genon:2013wba,Altmannshofer:2017fio}.
	\begin{figure}[!t]
	\begin{center}
	\includegraphics[height=2.75in]{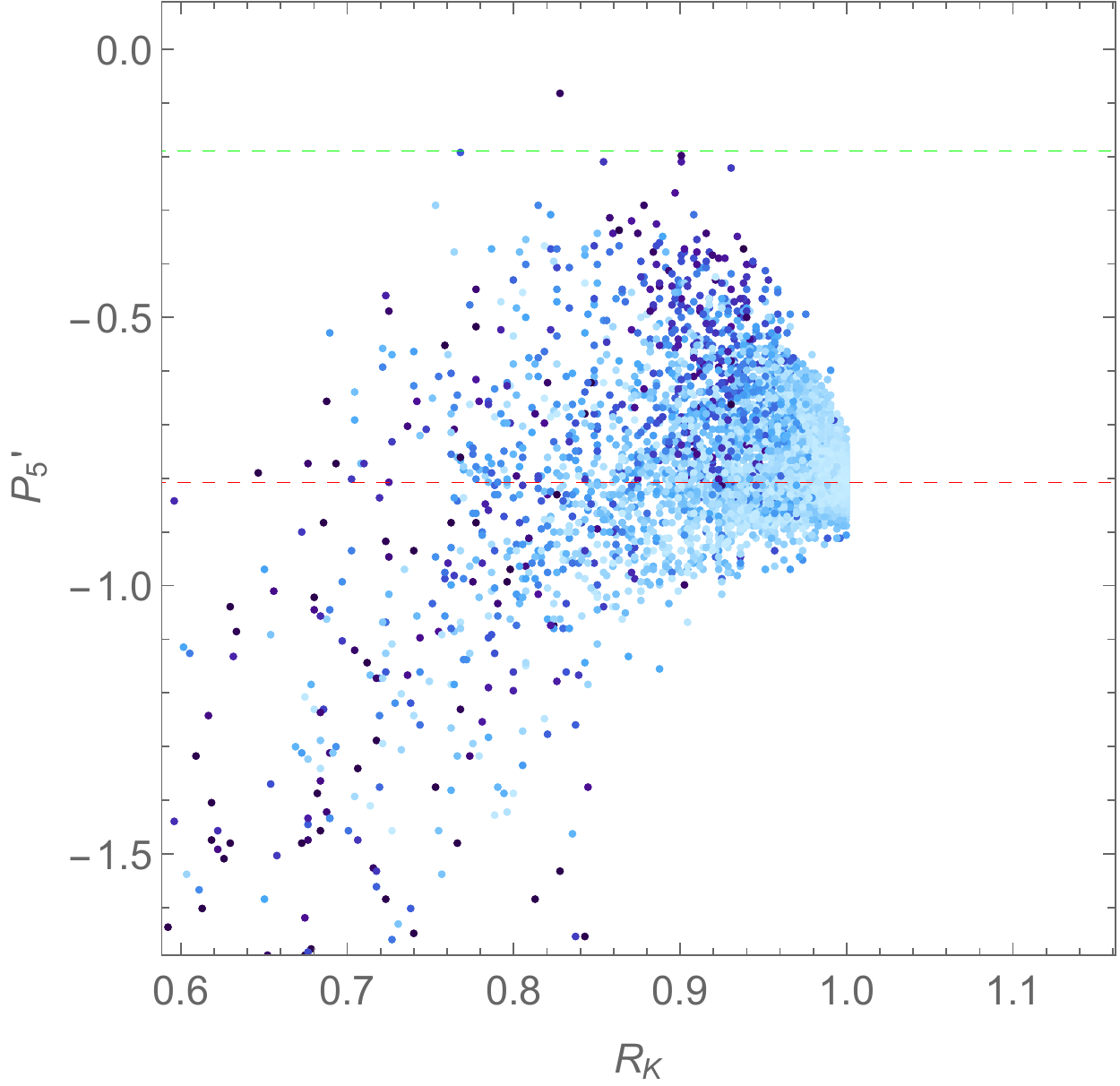} \raisebox{0.69cm}{\includegraphics[height=2.65in]{MBleg.pdf}}
	\caption{\label{fig:P5p} Correlation between $P_5^\prime$ and $R_K$, where the colors depict the
	size of the corrections to $\Delta M_{B_s}$. See text for details.}
	\end{center}
	\end{figure}
Our results for $P_5^\prime$ in the $q^2\in[4.3,8.68]\,{\rm GeV}^2$ bin are given in Figure \ref{fig:P5p}, where we plot the correlation between $P_5^\prime$ and $R_K$, visualizing again the size of the corrections to $\Delta M_{B_s}$
as different shades of blue. The SM prediction ${P_5^\prime}_{[4.3,8.68]}\approx -0.8$ and the experimentally
preferred value ${P_5^\prime}_{[4.3,8.68]}\approx -0.2$ \cite{Descotes-Genon:2013wba,Altmannshofer:2017fio}
are given as red and green dashed lines, respectively. 
From this final plot it is 
evident that the proposed composite model can address the $R_{K}$ and $R_{K^{\ast}}$ anomalies, being in agreement 
with $B_s - \bar B_s$ mixing and ${\cal B}(B_s \to \ell^+ \ell^-)$, while also the fit to the $P_5^\prime$ results 
can be improved (via the light-blue points approaching ${P_5^\prime}_{[4.3,8.68]}\approx -0.2$
and featuring $R_K \sim (0.7-0.8)$). It thus furnishes an interesting setup both with respect to the gauge hierarchy problem -
avoiding light top partners - and concerning the current pattern of experimental results in flavor physics.

\section{Conclusions}

We have scrutinized hints for lepton flavor universality violation in $B$-meson decays, focusing
first on the general properties of the anomalies in an EFT approach. Here, we emphasized a simultaneous
solution to the $R_K$ and $R_{K^\ast}$ anomalies via effects in right-handed lepton currents, not worked
out before. We stressed that this solution requires the dominant contribution originating from
electron (and not muon) currents and presented a composite Higgs scenario, where this pattern emerges
in a natural way. In fact, the model provides one of the very few scenarios, that features all ingredients
to consistently resolve the anomalies, without being actually constructed for that purpose: it {\it predicts} LFU violation
and sizable FCNCs involving the third quark generation, features a strong protection from FCNCs in the lepton sector, and
allows for the absence of ultra-light top partners at the LHC.
We also discussed the impact of operators addressing the $R_{K^{(\ast)}}$
measurements on other flavor observables, such as ${\cal B}(B_s \to \ell^+\ell^-)$, $\Delta M_{B_s}$,
and $P_5^\prime$. For the explicit model at hand, we found that all constraints are met, while
it is possible to simultaneously resolve the (more controversial) $P_5^\prime$ anomaly.

\section*{Acknowledgments}
We are grateful to Marco Nardecchia, David Straub and Jorge Martin Camalich, for useful comments and discussions.
FG thanks the CERN Theory Department, where part of this work was performed, for its hospitality.
The research of AC was supported by a Marie Skłodowska-Curie Individual Fellowship of the European Community’s Horizon 2020 Framework Programme for Research and Innovation under contract number 659239 (NP4theLHC14)
and by the Cluster of Excellence {\em Precision Physics, Fundamental Interactions and Structure of Matter\/} (PRISMA -- EXC 1098) and grant 05H12UME of the German Federal Ministry for Education and Research (BMBF).

\bibliographystyle{JHEP}
\bibliography{papers}

\end{document}